\documentclass[prl,nofootinbib,twocolumn,superscriptaddress,showpacs]{revtex4}

\usepackage[usenames,dvipsnames]{color}
\usepackage{graphicx}
\usepackage{amsmath}
\usepackage{amsfonts}
\usepackage{amssymb}
\usepackage{dsfont}
\usepackage{dcolumn}
\usepackage{bm}
\usepackage{slashed}
\usepackage{pstricks}

\usepackage{slashed}
\usepackage{multirow}
\usepackage{array}
\usepackage{rotating}
\usepackage{xcolor}
\usepackage{epstopdf}
\usepackage{fancybox}

\newcolumntype{x}[1]{
>{\centering}p{#1}}%

\newcommand{\GeV}      {~\mathrm{GeV}}

\def \cha{\widetilde{\chi}^{\pm}_1}

\newcommand{\beqn}{\begin{eqnarray}}
\newcommand{\eeqn}{\end{eqnarray}}
\newcommand{\be}{\begin{equation}}
\newcommand{\ee}{\end{equation}}

\newcommand{\mathsym}[1]{{}}

\def \cha{\tilde{\chi}^{\pm}_1}

\def \na{\tilde{\chi}^{0}_1}
\def \nb{\tilde{\chi}^{0}_2}

\def \n34{\tilde{\chi}^{0}_{3,4}}
\def \g{\tilde{g}}

\def \ta{\tilde{t}_1}
\def \tb{\tilde{t}_2}

\def\met100{\slashed{E}_T\geq 100 \GeV}

\newcommand{\st}{Stueckelberg~}

\usepackage{ulem,fancyvrb}

\pacs{95.35.+d, 12.60.Jv, 13.85.Rm, 98.70.Sa} 
\begin{document}

\title{ Higgsino dark matter model consistent with galactic\\ cosmic ray data
   and possibility of discovery at LHC-7}

\author{Ning Chen}
\affiliation{C.N.\ Yang Institute for Theoretical Physics, 
Stony Brook University, Stony Brook, New York 11794, USA}
  
\author{Daniel Feldman}
\affiliation{Michigan Center for Theoretical Physics,
University of Michigan, Ann Arbor, Michigan 48109, USA}

\author{Zuowei Liu}
\affiliation{C.N.\ Yang Institute for Theoretical Physics, 
Stony Brook University, Stony Brook, New York 11794, USA}

\author{Pran Nath}
\affiliation{Department of Physics, Northeastern University,
 Boston, Massachusetts 02115, USA}

\author{Gregory Peim}
\affiliation{Department of Physics, Northeastern University,
 Boston, Massachusetts 02115, USA}

\preprint{MCTP-09-yy; YITP-SB-10-34; NUB-TH-32zz}


\begin{abstract}
A solution to the PAMELA positron excess with Higgsino dark matter 
within extended supergravity grand unified (SUGRA) models is proposed.  
The models are compliant with the photon constraints  
recently set by Fermi-LAT and produce positron as well as 
 antiproton fluxes 
consistent with the PAMELA experiment.  The SUGRA models  considered have 
an extended hidden sector with extra degrees of freedom  which allow for a 
satisfaction of relic density consistent with WMAP. The Higgsino models are 
also consistent with the CDMS-II and XENON100 data and  are discoverable 
at LHC-7 with 1 fb$^{-1}$ of luminosity. The models are 
testable on several fronts.
\end{abstract}

\maketitle

\section{Introduction}  Recently, experiments detecting galactic cosmic rays have begun 
to probe the nature of the dark matter in the halo. The large excess observed of high 
energy positrons in the PAMELA experiment \cite{PAMELA} and the null results in 
the search for gamma ray lines with the Fermi-LAT experiment \cite{linesource} 
present a  challenge for particle theory. Some particle physics explanations have been 
proposed to explain the PAMELA data  consistent with the relic
abundance of dark matter including: a Breit-Wigner enhancement \cite{BW},
 a nonperturbative Sommerfeld enhancement \cite{somm,Feng}, and other 
possibilities \cite{FLNN,FLNP,Cohen,barger,adrs}. A nonthermal cosmological 
history is also a solution \cite{randallMoroi,DFGK,KLW}. Several astrophysics 
explanations have also been sought  \cite{astro}.
Within supersymmetry the positron excess can arise from the annihilation of neutralinos 
[lightest supersymmetric particle (LSP)] into $W^+W^-$ and/or $ZZ$. This comes about when the LSP is a pure wino 
\cite{randallMoroi,KLW},  a mixed wino-bino \cite{FLNN,Feldman:2010uv} or a Higgsino 
\cite{Olive:1989jg,Drees:1996pk,Baltz:1998xv,Kane:2001fz}. 
However, a wino LSP 
produces a large amount of monochromatic photons in its annihilation products which 
is edging close to the current upper limit set by the Fermi-LAT data \cite{Feldman:2010uv}. 

Here we present a supersymmetric model, which in contrast to other proposed
models, has mostly a Higgsino LSP  and can explain the relic abundance of
dark matter.   In addition,
we show that such a model fits the positron excess from PAMELA \cite{PAMELA} and 
is consistent with the antiproton flux, as well as with data from monochromatic photons 
that  arise via loop diagrams in the neutralino annihilation processes 
$\chi\chi\to\gamma\gamma,\gamma Z$ \cite{Bergstrom:1997fh}. 
We note that a bino-like LSP can also explain the PAMELA positron data when 
a substantial size  boost factor from the halo is allowed \cite{boosts}.

The monochromatic photon constraints from Fermi become very relevant when 
one tries to fit the PAMELA positron data via dark matter annihilations in the galactic halo as the cross 
section needed  to explain such data  is much larger than the naive estimation of the 
dark matter annihilation cross section from a thermal history. This expectation for the relic abundance
however can be modified which will be discussed. Such a modification can open new parameter
space in SUSY models where the relic density of dark matter is consistent with observations
and the flux of cosmics from dark matter at present temperatures can account for the  data.
This has implications for signatures of supersymmetry at the Large Hadron
Collider in the frameworks we discuss below.

 \section{Extended Abelian Models and Enhancement of Relic Abundance}
The simplest extension of the standard model (SM) which is gauge invariant, renormalizable, and
unitary  arises through a   \st mechanism \cite{korsnath,fln1}.  A $U(1)$ gauge boson $V_{\mu}$ gains mass $M$ through a  \st mechanism \cite{stueck} by directly absorbing an axion field
$\sigma$ through the 
combination $(M V_{\mu} +  \partial_{\mu}\sigma)^2$  which is gauge  invariant under the 
transformation $\delta V_{\mu} = \partial_{\mu} \lambda$, $\delta \sigma =-M \lambda$, 
and thus a transition to the unitary gauge produces a massive vector gauge boson without
the necessity of a Higgs mechanism. It is also well  known that the \st mechanism arises
quite naturally from a Green-Schwarz mechanism \cite{GS} with appropriate transformations.
Further, in reduction of higher  dimensional  theories the masses of the Kaluza-Klein states
arise from a \st mechanism and not from a Higgs mechanism. 
The \st mechanism
is indeed quite generic in  string theories (see e.g. 
\cite{Blumenhagen:2006ci}),  in extended supergravity theories, and in the compactification
of higher dimensional theories (for a review see \cite{pascos}). 

Interesting new physics arises if there is a hidden
sector  with minimally a $U(1)$ gauge field that mixes with the  hypercharge of the SM sector. 
A supersymmetric generalization of the \st mechanism leads to an extended
neutralino sector, i.e., where for each extra $U(1)_X$ factor one has two extra Majorana fields (Stinos)
which mix with the minimal supersymmetric standard model (MSSM) neutralinos.  
The above considerations generalize 
to  a set of  Abelian   $U(1)^n_X$  gauge groups, and 
such extensions lead to a mixing between fields in each
sector via gauge kinetic energy mixings and mass mixings. 

We implement this extension to study a class of supergravity unified models which 
 allow the possibility of explaining the PAMELA data without recourse to large 
 clump  factors in the halo of the Galaxy. We uncover a new situation where the 
 LSP is actually a nearly pure Higgsino under
 radiative electroweak symmetry breaking with mass in the range $\sim (110-190)~ \rm GeV$
 with the hidden sector components of the LSP  being suppressed.

 Thus we consider a supergravity grand unified 
model \cite{msugra, NUSUGRA} having an extra hidden sector with a product gauge 
group $U(1)^n_X$ \cite{FLNN} which mixes with the hypercharge via mass 
terms generated by the \st mechanism and without loss of generality
via gauge kinetic mixing. For simplicity, we give a summary for  the case of a single $U(1)_X$,  and
the generalization for a product gauge group follows analogously. In  the 
 vector sector the mass mixing and gauge kinetic energy  mixing  is of the form
$-2 M_X M_Y X^{\mu} Y_{\mu} -(\delta/2) X^{\mu \nu} Y_{\mu \nu}$ and
in the 
neutralino sector the mass mixing is of the form 
$\psi_{\rm st} (M_X \lambda_X +M_Y \lambda_Y) + h.c.$ 
while the kinetic mixing leads to
$-i\delta(\lambda_X \sigma \cdot \partial \bar \lambda_Y+ (Y\leftrightarrow X) )$, 
where $X$ denotes the hidden sector $U(1)$ and $Y$ is the hypercharge of the MSSM;     
$\psi_{\rm st}$ is a fermonic field  that arises out of a chiral \st  supermultiplet  
  and  $M_Y: M_X$ and $\delta$ are small, i.e. on the order of $10^{-2}$ or smaller \cite{fln1}. 
  Such  additional states remain in contact with the thermal bath prior to freezeout in the early universe.  In the absence
  of hidden sector soft masses, a direct study of the mass matrix in the neutralino sector gives rise to   
   a  mass degeneracy $g_{\rm hid}$  for  the hidden sector    neutralinos with the LSP, which 
in turn,   can have   a degeneracy $g_{\rm vis}$ with other visible sector sparticles    \cite{fln2}. 
  Coannihilations can then produce an  enhancement  of the relic density by a factor 
$f_E$ \cite{FLNN} so that 
\be
\Omega_{\tilde \chi^0} h^2\simeq ~f_{E}\times\Omega_{\tilde \chi^0}^{\rm MSSM}h^2, ~~~~
f_E= \bigg[1 + \frac{g_{\rm hid}}{g_{\rm vis}}\bigg]^2. 
\ee
Generalizing to the case of a $U(1)^n_X$ extended hidden sector $g_{\rm hid}= 2n$, 
and thus for the case  $g_{\rm vis}=1$, one finds $f_E= (2n+1)^2$ which gives $f_E= 25 (49)$ for 
$n=2(3)$.  In this  extended model the neutralino mass matrix will be $(4+2n)\times(4+2n)$ 
dimensional. We assume that the LSP lies in the visible (MSSM) sector. Because of coannihilations
in the visible sector, the full enhancement is never achieved, however one finds large enhancements 
of size $(10-20)$ or larger with a degenerate hidden sector and only $2-3$ additional $U(1)$s 
which is  sufficient for compatibility with  the WMAP constraint \cite{WMAP} since the models 
considered have the relic density in the range $\sim(2-6)\times10^{-3}$ if there were no Abelian hidden sector.

 \section{Low Mass Higgsino LSP in extended SUGRA and Fermi Photons}
We discuss now the details of the Higgsino-like neutralino  models. 
Since the extra weak mixing discussed
above is small, it has negligible effects on the soft parameters
at the weak scale.
 The model
parameters that dictate annihilation cross sections can then be described by the input parameters 
given in Table(\ref{tab1}). The models (P1-P3)  listed in Table(\ref{tab1}) have a neutralino that 
is  dominantly a Higgsino,  with about $2\%$ remaining in the gaugino content. For comparison, we 
also exhibit  the mixed wino-bino  (WB) model \cite{FLNN}  which has a significant  wino content $\sim 49\%$
 of the total eigencontent along with a comparable bino content.  All four models satisfy the current 
experimental constraints from flavor physics and limits on sparticle masses (see e.g.~\cite{chen}). Their neutralino masses lie in the
range (110-190) GeV and have a spin independent cross section of size (5-10)$\times10^{-45}\rm cm^2$  
consistent with the upper bounds from the CDMS-II and XENON100 \cite{Aprile:2010um}. Further, 
some of the models possess several rather light sparticles in their  spectra, namely the  charginos, 
neutralinos, gluino and in some cases the stop, and are thus good candidates for discovery at the LHC. 

 \begin{table}[t!]
\begin{tabular}{c|cccccc|cccc} 
\hline
Model 	& $m_0$	& $M_1$	& $M_2$ & $M_3$& $A_0$	& $\tan\beta$ 
 & $\mu'$ & ${M}^{'}_1$ & ${M}^{'}_2$& ${M}^{'}_3$    \\
\hline
P1 & 1033 & 1600 & 1051 & 120 & 2058 & 13 &  195  & 683 &836  & 259 \\
P2 &1150 & 1600 &1080 & 160 & 2080 & 15 &  152  & 684 &859  & 347\\
P3 & 950  & 1425& 1820& 748 & 1925 & 25 &  109  & 617 & 1453  & 1589\\
WB & 2000 &400 &210& 200& 300 & 5  &  562  & 170 &163 & 441 \\
\end{tabular}
\caption{Parameters which produce an LSP which  are mostly Higgsino (P1-P3), or mixed 
wino-bino, WB. Here $m_0 (A_0)$ is the universal scalar mass (trilinear coupling), 
$M_1, M_2, M_3$ are the gaugino masses  at the GUT scale for the gauge groups $U(1)_Y, SU(2)_L, SU(3)_C$ 
and $\tan\beta$ is the ratio of the two Higgs vacuum expectation values in the MSSM. The parameters that enter the neutralino mass 
matrix  at  scale 
$Q =\sqrt{M_{\ta} M_{\tb}}$  are $(\mu',{M}^{'}_1,{M}^{'}_2,{M}^{'}_3)$,  where $\mu'$ is the 
Higgs mixing parameter. 
The models have also been run through both SuSpect and SOFTSUSY via micrOMEGAs 
\cite{susypackage}.  Here
$m_{\rm top}^{\rm pole}$=173.1 GeV.}
\label{tab1}
\end{table}
 \begin{table}[t!]
\begin{center}
\begin{tabular}{cccc|cc}
$E_\gamma$    &  Einasto & NFW & Isothermal  & Model &   
$\langle \sigma v\rangle_{\gamma Z,[\gamma\gamma]}^{\rm theory}$\\
\hline
180[190]   &  4.4[2.3] &  6.1[3.2] & 10.4[5.5] &  P1 & 0.24[0.08] \\
130[150]  &  5.3[2.5] &  7.3[3.5]&  12.6[6.0] & P2  & 0.23[0.09]  \\
90[110]  &  4.3[0.7] &  6.0[1.0]  & 10.3[1.7] & P3 & 0.18[0.09]\\ 
150[160] &  5.9[2.0] &  8.2[2.7]&  14.1[4.7] &  WB  &  7.00[1.29]  \\ 
\end{tabular}
\caption{Cross sections $\langle \sigma v\rangle_{\gamma Z}$ 
and $\langle \sigma v\rangle_{\gamma\gamma}$ upper limits 
($10^{-27}$cm$^3$/s)  \cite{linesource}
for 3 halo profiles (Einasto, Navarro-Frenk-White (NFW), and Isothermal) along with  predictions for (P1-P3)  and WB. 
The mostly Higgsino models (P1-P3) are unconstrained
by any profile  while the  mixed wino-bino model, WB, is  on the edge.}
\label{tab2}
\end{center}
\end{table} 
In Table(\ref{tab2}) we give the theoretical predictions of the Higgsino LSP models for the $\gamma Z$ 
and $\gamma\gamma$ modes and exhibit the current upper limits from the Fermi-LAT search 
for photon lines using three different halo profiles. One finds that the theoretical predictions for 
the Higgsino models (P1-P3), are well below the current upper bounds from Fermi-LAT, 
by about a factor of 10, for the most restrictive profile, while the mixed wino-bino  model, WB,
is close to the edge of the limits.
There are sources of photons arising from  bremsstrahlung 
that could mimic the line signature of monochromatic photons.
The contributions from bremsstrahlung to the line 
signals can be significant or even dominant over the ones from the loop processes 
\cite{Bergstrom:2005ss}.  However, the additive effects from 
bremsstrahlung to the line source are small for the models considered 
here which have the dark matter in the mass range $\sim (110-190)~\rm GeV$. 
This is due to the fact that the maximal energy the photon can carry is 
$E_{\gamma}^{\rm max} = M_{\chi}(1-M_W^2/M^2_{\chi})$ in the process 
$\chi\chi\to WW\gamma$, and the energy of the monochromatic photons 
via $\chi\chi \to \gamma X$  is 
$E_{\gamma}=M_{\chi}[1-M^2_X/(4M^2_{\chi})]$. Thus the photons arising from 
the process $\chi\chi\to WW\gamma$ have energy whose location in the energy spectrum is at least 
$\sim (23-40)~\rm GeV$  below the monochromatic photons in $\gamma Z, \gamma\gamma$ 
final states for dark matter mass in the  range $\sim (110-190)~\rm GeV$. 
For a related discussion see \cite{mario}.

The Higgsino  models typically have a small $\mu$ and large $m_0$ and lie on the boundary of the 
radiative electroweak symmetry breaking curve, i.e., the Hyperbolic Branch \cite{ArnowittNath,chatto}. 
It is the smallness of $\mu$ relative to the soft gaugino masses that makes the three lightest particles,
the two lightest neutralinos and the lighter chargino,  essentially degenerate in mass \cite{chatto}.
In this region $\mu$ (and some of the sparticle spectrum) is very sensitive to small changes in 
the input parameters at the GUT scale. On the other hand since $\mu$ is small, one is in a less fine 
tuned region. Alternately, instead of working down from the high scale, one could simply generate these
Higgsino-like LSPs  directly by inputs at the weak scale. We have checked this for the models discussed here.
This is evident from Table(\ref{tab1}).

  \begin{figure*}[t!]
   \begin{center}
  \includegraphics[width=8.2cm,height=6.25cm]{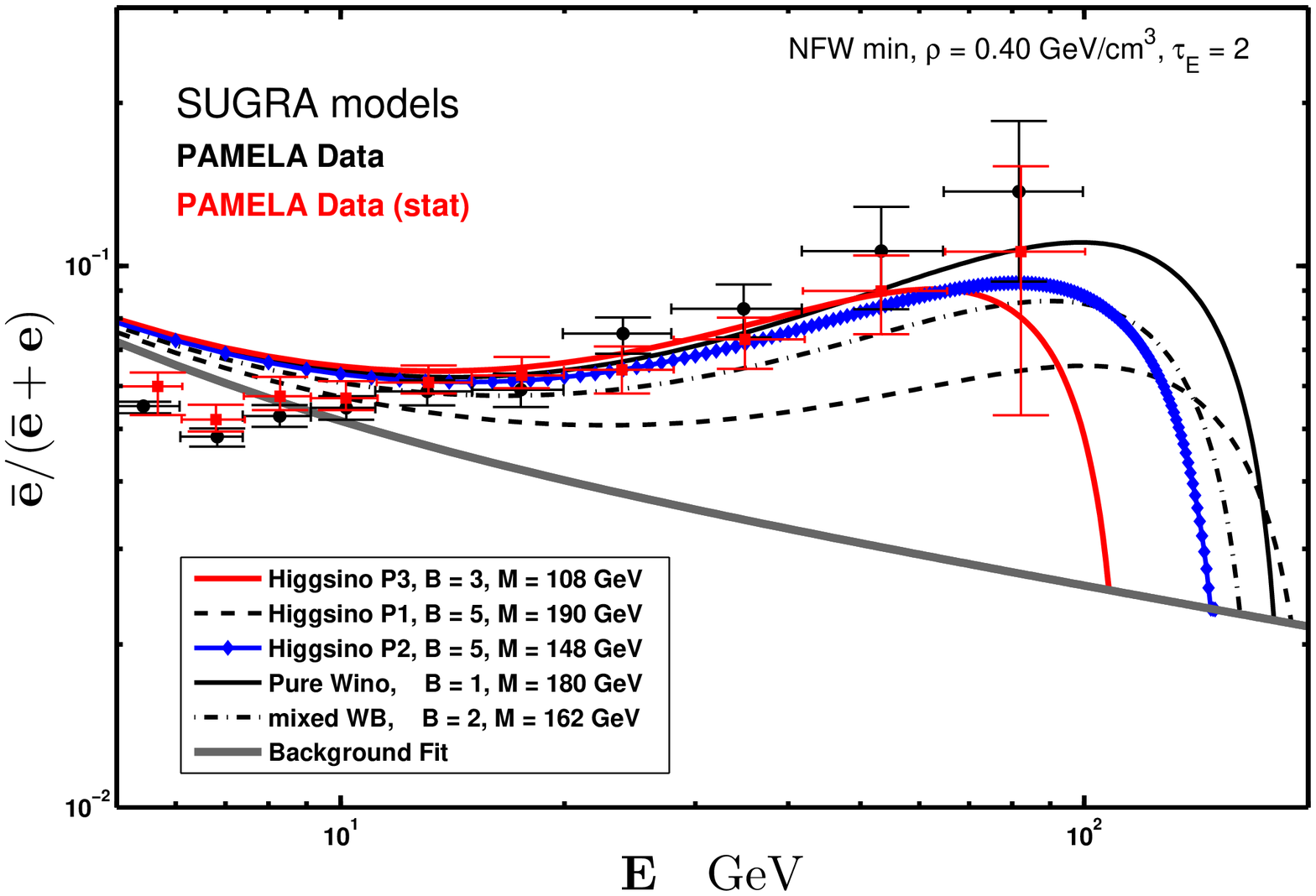}\hspace{1cm} 
  \includegraphics[width=8.3cm,height=6.25cm]{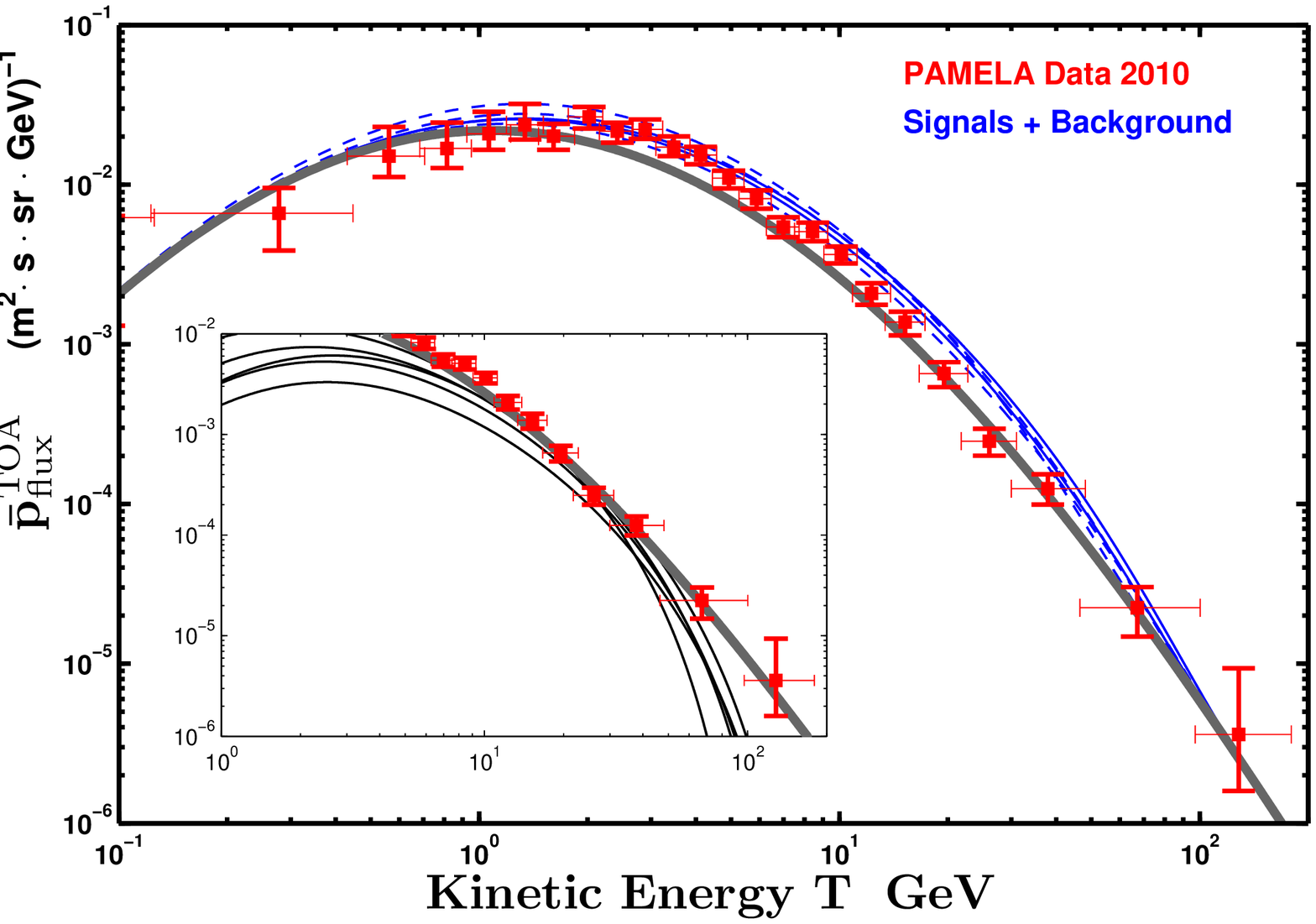}\\ 
     \includegraphics[width=7.5cm,height=6.9cm]{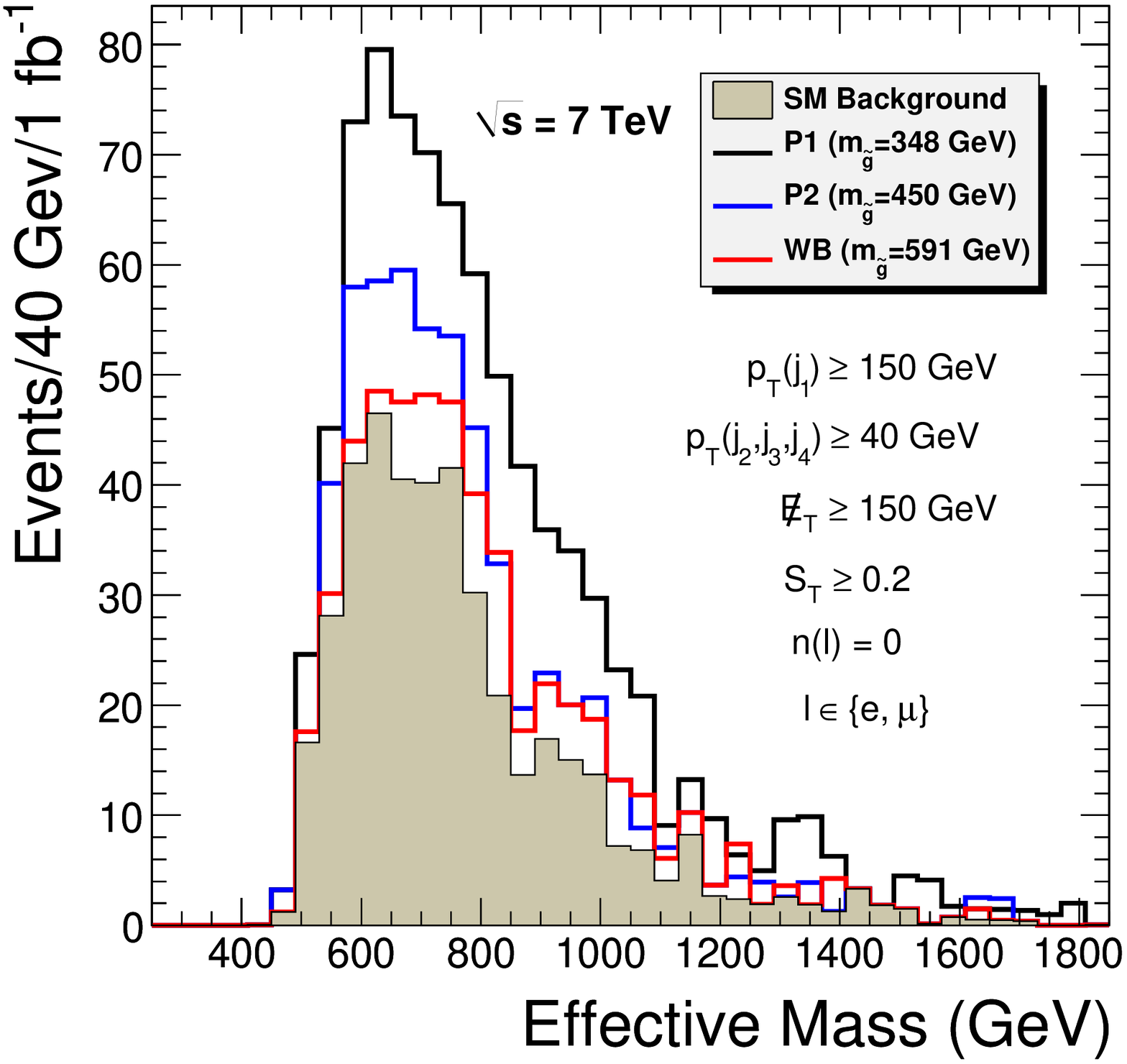} \hspace{1.1cm} 
    \includegraphics[width=7.5cm,height=6.9cm]{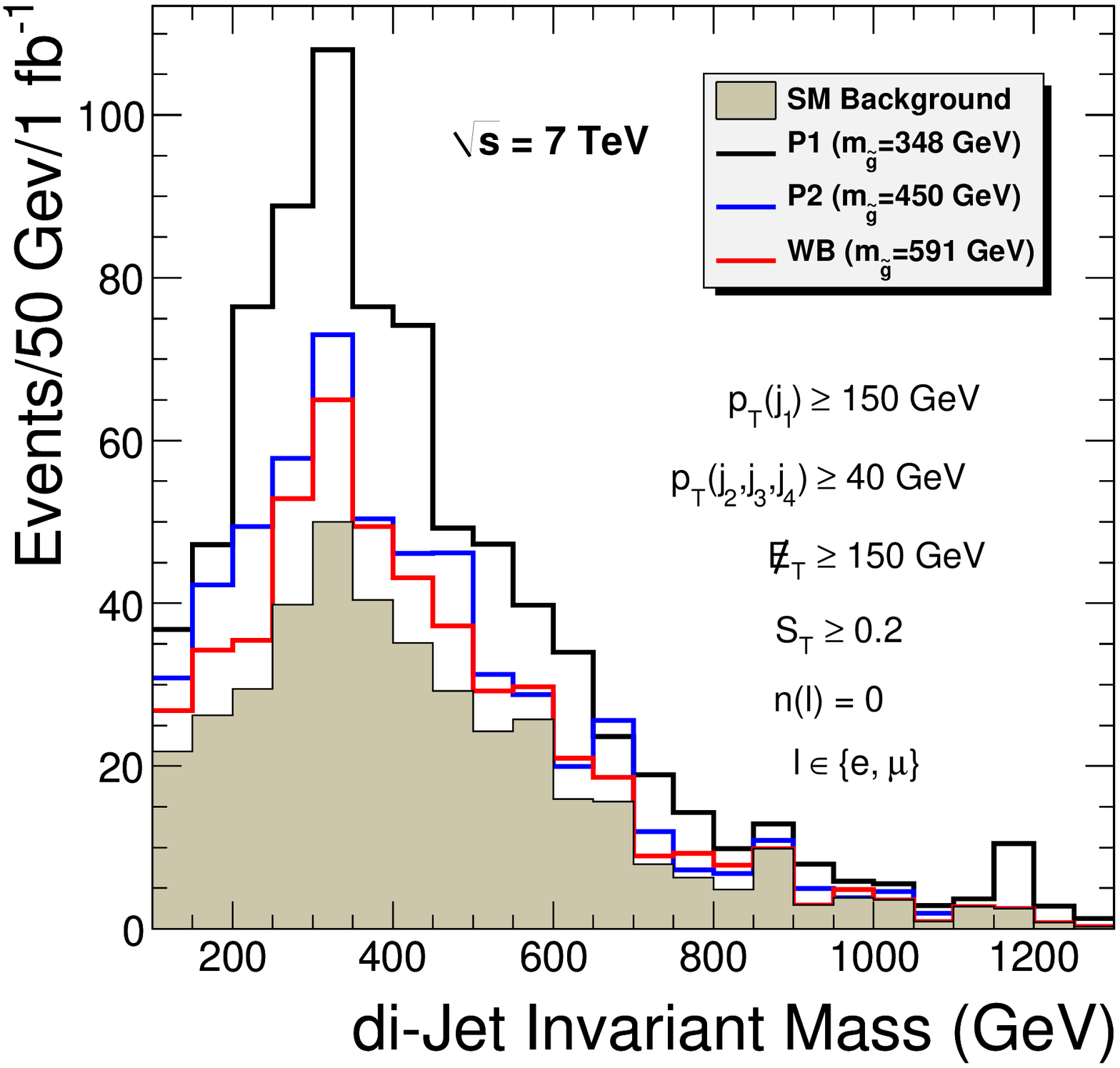}  
\caption{(Color online) Top left:  PAMELA positron excess  and the Higgsino models (P1-P3). 
The  wino dominated model is also shown for comparison along with a mixed wino-bino model, WB. 
Top right:   The PAMELA  $\bar p$  flux and the  predictions are seen to be compatible with the data. 
Equal dark matter densities and boosts are taken in both top panels.  
Lower left: SUSY plus background events vs effective mass at 1~fb$^{-1}$ for the signature cuts
shown in the panel for models P1, P2 and WB. Lower right: SUSY plus background events vs 
the di-jet invariant mass  at 1~fb$^{-1}$ for signature cuts shown in the panel for the models 
P1, P2 and WB.  Both lower plots do not show model P3 due to its suppressed cross section at LHC-7.
The legends labeling the model names (not colors) indicate the model class
in all figures.}
\label{fig1}
  \end{center}
\end{figure*}

\section{~~~Positrons from Higgsinos\newline and Mixed Winos}
Next, we discuss the positron excess prediction in the Higgsino-like model. In Higgsino and wino models, 
the high energy positron flux can arise from $WW$ and $ZZ$ production from the neutralino 
annihilation in the halo with approximate cross sections at leading order  \cite{Olive:1989jg}
\begin{equation}
\langle \sigma v\rangle(\chi\chi\to VV) \simeq 
\frac{g_2^4}{C_V 2\pi M^2_{\chi}} \frac{(1-x_V)^{3/2}}{(2-x_V)^2}, 
\label{approx}
\end{equation}
where $V=(W,Z)$, $x_V=M^2_V/M^2_{\chi}$,  $C_W=16~(1)$ for Higgsino (wino) models and 
the $ZZ$ production is only significant for Higgsino models where $C_Z=32\cos^4(\theta_W)$.
For the models (P1-P3)  the LSP is mostly a Higgsino with only a very small portion being gaugino.  
Here the cross sections that enter in the positron excess are  size 
$\langle \sigma v\rangle(\chi\chi\to WW,ZZ)_{\rm Higgsino}\lesssim 4 \times 10^{-25}\rm cm^3/s $.

The positron flux from the Higgsino dark matter can be described semianalytically
(for early work, see  \cite{Baltz:1998xv}). The flux  enters as a solution to the diffusion loss
equation, which is solved in a region with a cylindrical boundary. The particle physics 
depends on $\langle \sigma v \rangle_{\rm halo} $, and $dN/dE$, the  fragmentation functions / energy distributions \cite{Baltz:1998xv}. The astrophysics depends on the dark matter profile  \cite{prof}, 
and on the energy loss in the flux from the presence of  magnetic fields and from scattering off 
galactic photons.  A boost factor which parametrizes the 
possible local inhomogeneities of the dark matter distribution can be present.  
Recent results from N-body simulations indicate that large dark matter 
clumps within the halo are unlikely \cite{Brun:2009aj} \cite{Kamionkowski:2010mi}. 
The boost $B$ we consider here is small, as low as $\sim (2-3)$. 
The background taken is consistent with the GALPROP \cite{Moskalenko:1997gh} 
model generated in Ref. 1 of \cite{KLW}. The antiproton flux follows rather analogously  
(for an overview and some fits see e.g.  \cite{Cirelli08}). In this analysis  
the antiproton backgrounds are consistent with  \cite{Bringmann:2006im}, and the results 
for the pure wino case considered are consistent with  \cite{KLW,FLNN,Feldman:2010uv}.

The full analysis is exhibited in the  upper left panel of Fig.(\ref{fig1}) where we show  fits 
to the PAMELA positron fraction   \cite{PAMELA}. For comparison we also show the 
essentially pure wino case, which requires no boost (clump),   but as mentioned in the introduction,
will generally lead to an overproduction of  photons. Model P3 requires a boost of only 
$\sim (2-3)$ as the LSP is light, $\sim 110~{\rm GeV}$. For this case, the $\bar p$ 
flux is slightly larger at lower kinetic energy, but still consistent with the data. A pure 
wino at 110 GeV  would give a cross section about 10 times larger
relative to the Higgsino model at 110 GeV. Including the boost factor of 3 for the 
Higgsino model, the pure wino is then $(3-4)$ times stronger in its flux, 
and this is another reason  a pure wino at 110 GeV would fail - it would overproduce the antiprotons, 
whereas the Higgsino with minimal boost is consistent. Thus,  in  the upper  right panel of 
Fig.(\ref{fig1}) we  give a comparison of the $\bar p $ flux with the recently released data 
from Ref. 3 of  \cite{PAMELA}. Indeed it is seen that the theoretical prediction of the
 $\bar p$ flux is in perfectly good accord with this  data. 
 We note there are other processes beyond the leading order  that could produce SM 
gauge boson final states.  For diboson final states, these corrections are
rather small and lead to a small shift downward in the clump factor used (see Ref. 2 of \cite{Ciafaloni:2009tf}).
 The minimal boost  utilized here is rather different compared to those in analyses
of  bino-like LSPs  which use boosts of size $10^2$ or larger \cite{boosts} to fit the data. 
The analysis we present does not attempt to explain the high energy $e+ \bar e$ data  
\cite{LR,Meade:2009iu}. This could be explained with an 
additional electron source\cite{KLW}.

\section{Signature Analysis: LHC, $\sqrt s=7$ TeV}
As mentioned above, some of the colored sparticles in the Higgsino-like models are rather light which 
is encouraging for possible early  discovery of this class of models at $\rm LHC-7$ \cite{Nath:2010zj}. To 
achieve a significance necessary for discovery, i.e., $S\geq\max\{5\sqrt{B},10\}$, it is essential to 
have a reliable SM background computation. In our analysis we simulate the SM backgrounds 
\cite{Peim} using MadGraph 4.4  \cite{Alwall:2007st} for  parton level processes, PYTHIA~6.4 
for  hadronization and PGS-4 for detector simulation \cite{pythiapgs}. The $b$-tagging efficiency 
in PGS-4 {is} based on the technical design reports of CMS and ATLAS \cite{CMSAtlas} (see \cite{Peim}).  The sparticle spectrum and branching ratios for the signal analysis is 
generated using computational packages for supersymmetric models  \cite{susypackage}.   

The models we consider for the LHC-7 analysis have rather light gluinos in the mass
range $(\sim 350 - 600)~\rm GeV$ (see also \cite{Chen:2010kq}).  The production cross sections for these models are dominated by gluino production
and the branching fractions are dominated by either the radiative decay of the
gluino $\g \to g \na, \g \to g \nb $ (Higgsino-like model P1) or a combination of the radiative decays above and the
three body decays $\g \to \cha (b \bar t +h.c.)$ (Higgsino-like model P2) or effectively just the 3 body decays
producing both $\cha$ and $\nb$ with substantial rates (mixed wino-bino model WB) .
The subsequent decays follow from the chargino and neutralino into standard model quarks and leptons. Decays
into the degenerate hidden sector particles near the LSP mass are suppressed.

In the lower left panel of Fig.(\ref{fig1}) we give an analysis for the Higgsino models P1 and P2 with 
the number of SUSY events in 40 GeV bins at  ${\rm 1~fb^{-1}}$ of integrated luminosity vs the 
effective mass  defined to be the sum of the $p_T$ of the four hardest jets plus missing energy. The 
cuts used are exhibited in the panel. In the lower right panel of Fig.(\ref{fig1}) we give an analysis 
for the Higgsino models P1 and P2 with the number of SUSY events in 50 GeV bins at 
${\rm 1~fb^{-1}}$ of integrated luminosity vs  the di-jet invariant mass   where the cuts used are 
exhibited in the panel. For comparison we also give an analysis of the mixed wino-bino model, WB, 
in both lower left and lower right panels. Since the gluino is relatively light and the squarks are 
heavier, the 3 body decays of the gluino  dominate resulting in rich di-jet signals and effective mass. 
We note that while the model P3 provides a good fit to the PAMELA data  and its 
photon flux is an order of magnitude below the current limits,  it has a heavy  ($\sim$ 1.5 TeV) gluino 
and  would  not produce an identifiable 
signal in the early LHC data.

 \section{Conclusion}
We have presented here a solution to the PAMELA data and the  Fermi photon data
with a Higgsino-like LSP which can also be made compliant with WMAP.  It is shown 
that the models considered are consistent with the current very stringent limits on 
$\gamma\gamma$ and $\gamma Z$ production from Fermi-LAT which put the pure wino LSP 
models close to the edge of the upper limit of experiment. Further, the Higgsino LSP 
models are consistent with the upper limit  from the XENON100 experiment 
and  will be testable  in improved dark matter experiments. We find that  LHC-7 can 
realistically  probe these models up to gluino masses of  $\sim 600$ GeV with 
1 fb$^{-1}$ of data.  However, one would need  
larger  integrated luminosity to carry out precise mass reconstructions. The above 
presents an interesting possibility of having a  low mass gluino from the radiative breaking 
of the electroweak symmetry which can be produced at the LHC in early runs and also 
having a mostly Higgsino LSP giving rise to PAMELA positron excess. Thus the class 
of models discussed here can be tested on multiple fronts.


\section{Acknowledgements}  
This research is  supported in part by Department of Energy (DOE) Grant No.  DE-FG02-95ER40899,  and 
the U.S. National Science Foundation (NSF) Grants No.
 PHY-0653342, No. PHY-0704067, and  No. PHY-0757959,
 and in addition by the NSF
 through TeraGrid resources provided by National Center for Supercomputing Applications (NCSA), 
 Texas Advanced Computing Center (TACC), Purdue University and Louisiana Optical Network Initiative (LONI) under Grant No. TG-PHY100036.


\end{document}